\documentstyle[aps,twocolumn,graphicx]{revtex}

\newcommand{\myscalebox}[1]{\scalebox{0.4}[0.45]{#1}}
\newcommand{\myscaleboxb}[1]{\scalebox{0.35}[0.4]{#1}}
\newcommand{\myscaleboxc}[1]{\scalebox{0.5}[0.5]{#1}}
\newcommand{\captionA}
{Average ground state energy $e$ per spin 
as function of AF-bond concentration $p$ for
system sizes $L=3,4,6,14$. Lines are guides for the eyes only.}

\newcommand{\captionB}
{Binder cumulant of magnetization as function of AF-bond concentration $p$
for system size $L=3,5,8,10,14$. Two typical error bars are given. 
Lines are guides for the eyes only.}

\newcommand{\captionC}
{Scaled plot of Binder cumulant. Line is guide for the eyes only.}

\newcommand{\captionD}
{Scaled plot of magnetization. The inset show the
raw data for $L=3,5,14$.  Lines are guides for the eyes only.}

\newcommand{\captionE}
{Average overlap $\langle |q| \rangle$ value as function of 
AF-bond concentration $p$ for $L=3,4,8,14$. 
With increasing concentration
more and more spins belong to clusters which contribute to the
degeneracy of the ground state, so the average overlap value
decreases. Where the onset of the spin glass behavior is
located exactly cannot be seen from this figure. 
Lines are guides for the eyes only.}

\newcommand{\captionF}
{Distribution $P(|q|)$ of overlaps for AF-bond 
concentrations about $p\approx 0.18$ for $L=3,4,8,14$. 
A finite fraction of spins
is contained in small clusters which can take two orientations in
the ground state. For $L\to\infty$ a delta-function is obtained similar
to the distribution found for the ground states of random-field Ising
systems. Lines are guides for the eyes only.}

\newcommand{\captionG}
{Distribution $P(|q|)$ of overlaps for AF-bond 
concentrations about $p\approx 0.23$ for $L=3,4,8,14$. 
The distribution is broad and extends to $q=0$ for
all sizes $L$, indicating a spin-glass behavior. 
Lines are guides for the eyes only.}

\newcommand{\captionH}
{Average variance $\sigma^2(|q|)$ of overlap distribution 
as function of AF-bond 
concentration $p$. For small concentrations  the width of
the distribution shrinks to zero with increasing size $L$. For larger
values the spin-glass phase is characterized by broad distributions
of overlaps. Lines are guides for the eyes only.}

\newcommand{\captionI}
{Average fraction $X_{0.5}$ of overlap distribution below $q=0.5$
as function of AF-bond concentration $p$ for sizes $L=3,4,8,14$. 
In the spin-glass phase the
distribution of overlaps extends to $q=0$ for all lattice sizes.
Lines are guides for the eyes only.}

\newcommand{\captionK}
{Example of the Cluster-Exact Approximation method. A part of a spin glass
is shown. The circles represent lattice sites/spins. Straight lines represent
ferromagnetic bonds the jagged lines antiferromagnetic interactions. The
top part shows the initial situation. 
The construction starts with the spin at the center. The bottom part 
displays the final stage.
The spins which belong to the cluster carry a plus or minus sign which
indicates how each spin is transformed, so that only ferromagnetic
interactions remain inside the cluster. All other spins cannot be added
to the cluster because it is not possible to multiply them by $\pm 1$
to make all adjacent bonds positive. Please note that many other combinations
of spins can be used to build a cluster without frustration.}

\newcommand{\figA}{
\begin{figure}[htb]
\begin{center}
\myscalebox{\includegraphics{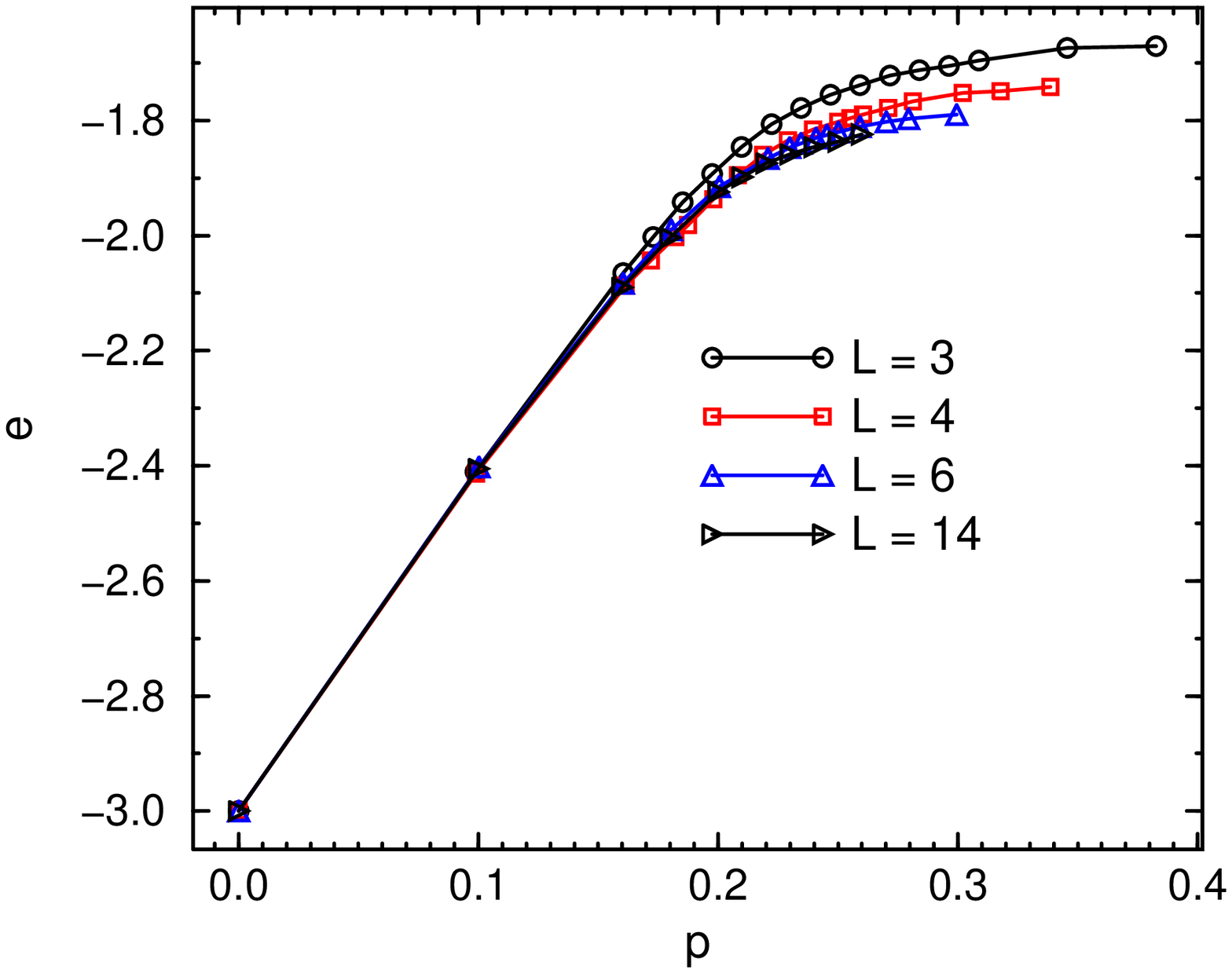}}
\end{center}
\caption{\captionA}
\label{fig_energy}
\end{figure}
}

\newcommand{\figB}{
\begin{figure}[htb]
\begin{center}
\myscalebox{\includegraphics{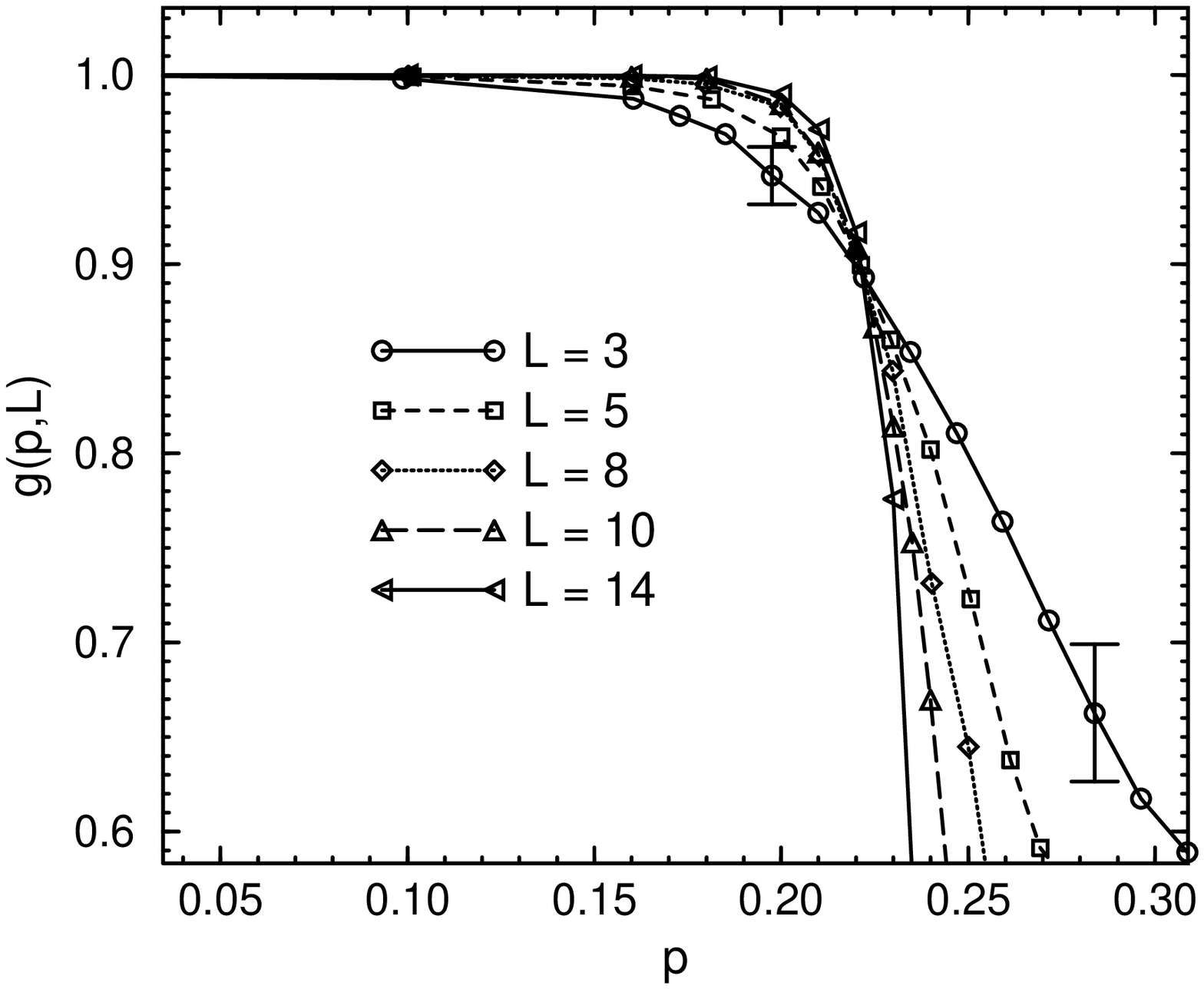}}
\end{center}
\caption{\captionB}
\label{fig_binder}
\end{figure}
}

\newcommand{\figC}{
\begin{figure}[htb]
\begin{center}
\myscalebox{\includegraphics{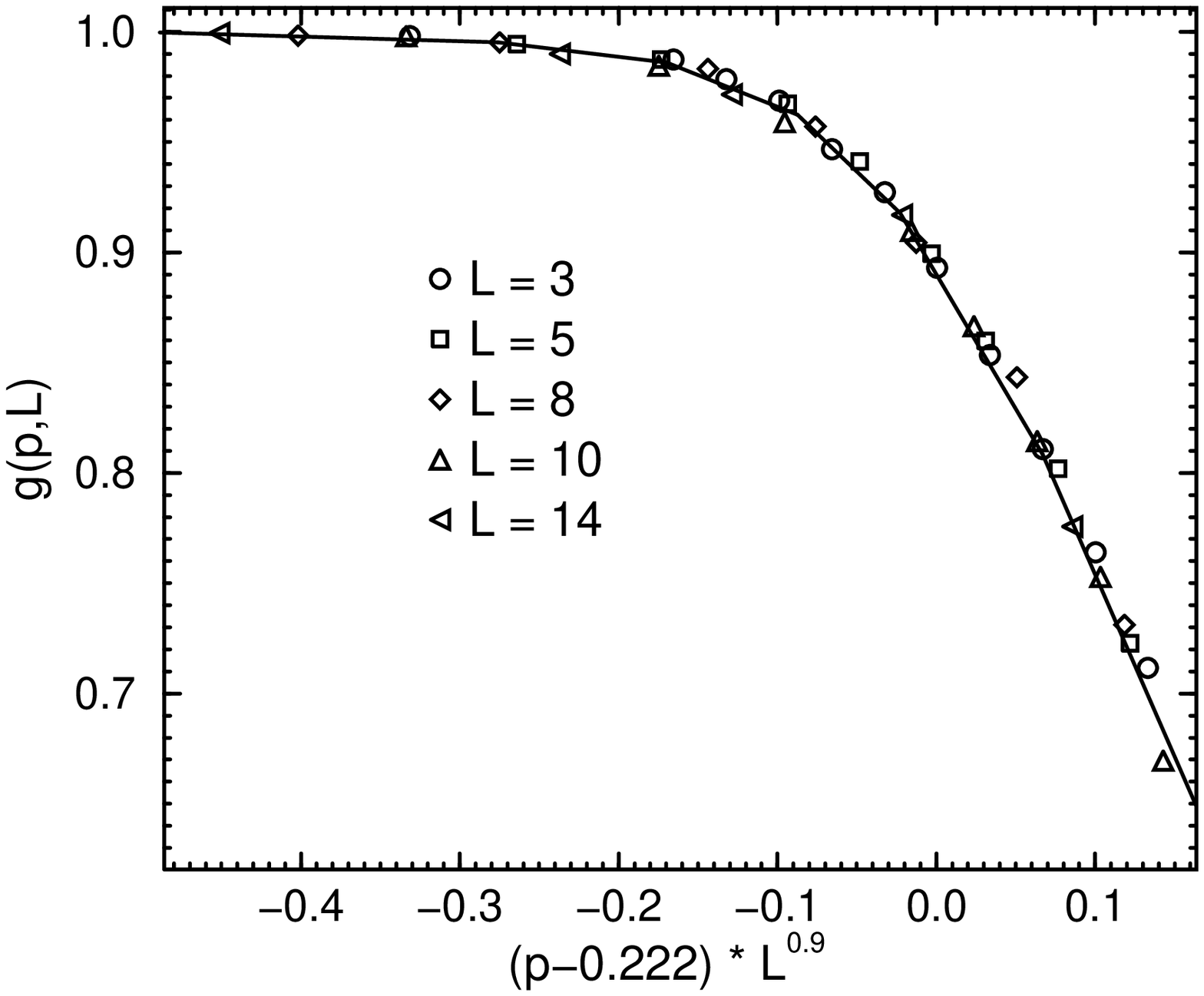}}
\end{center}
\caption{\captionC}
\label{fig_binder_scale}
\end{figure}
}

\newcommand{\figD}{
\begin{figure}[htb]
\begin{center}
\myscaleboxb{\includegraphics{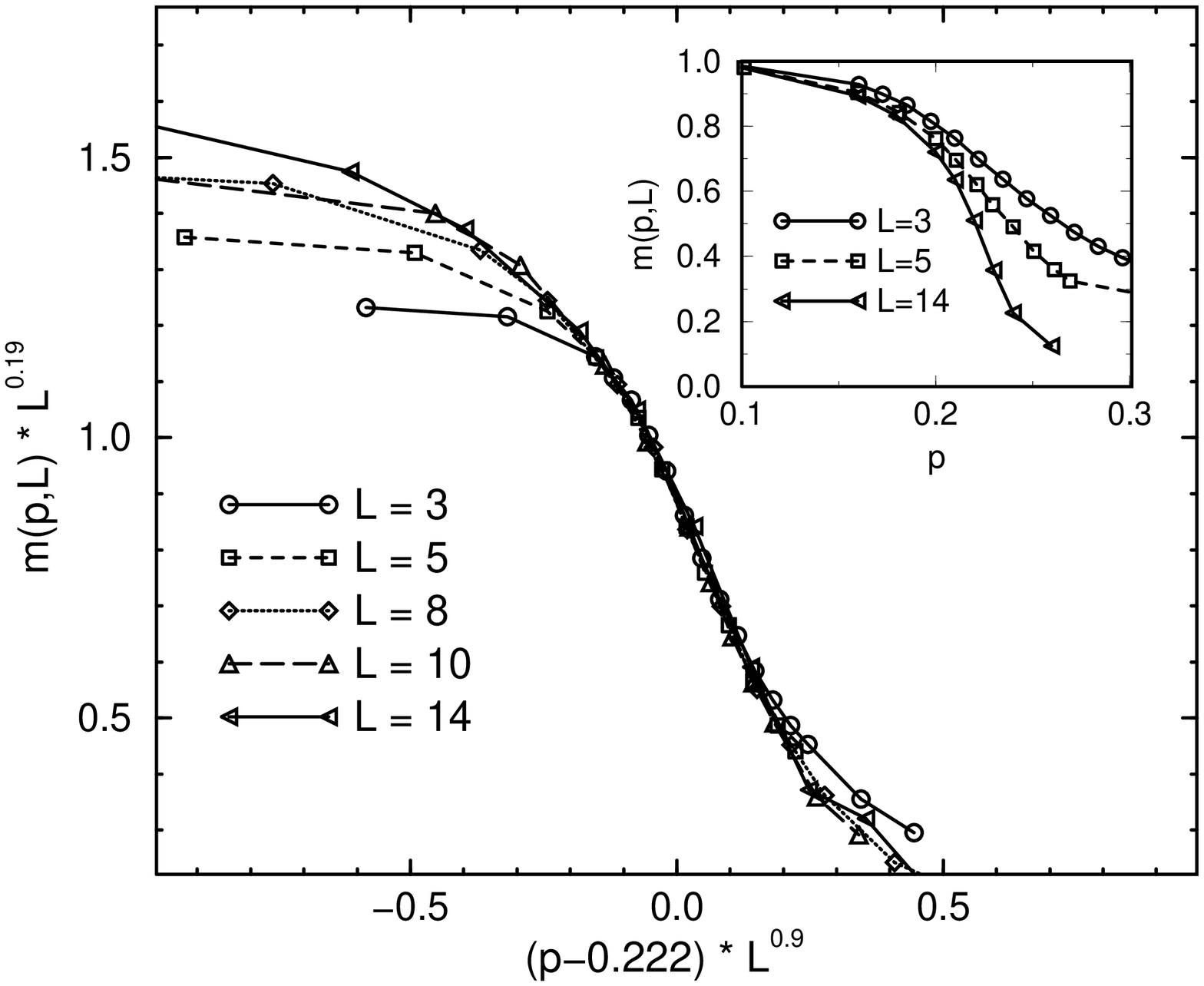}}
\end{center}
\caption{\captionD}
\label{fig_magn_scale}
\end{figure}
}

\newcommand{\figE}{
\begin{figure}[htb]
\begin{center}
\myscalebox{\includegraphics{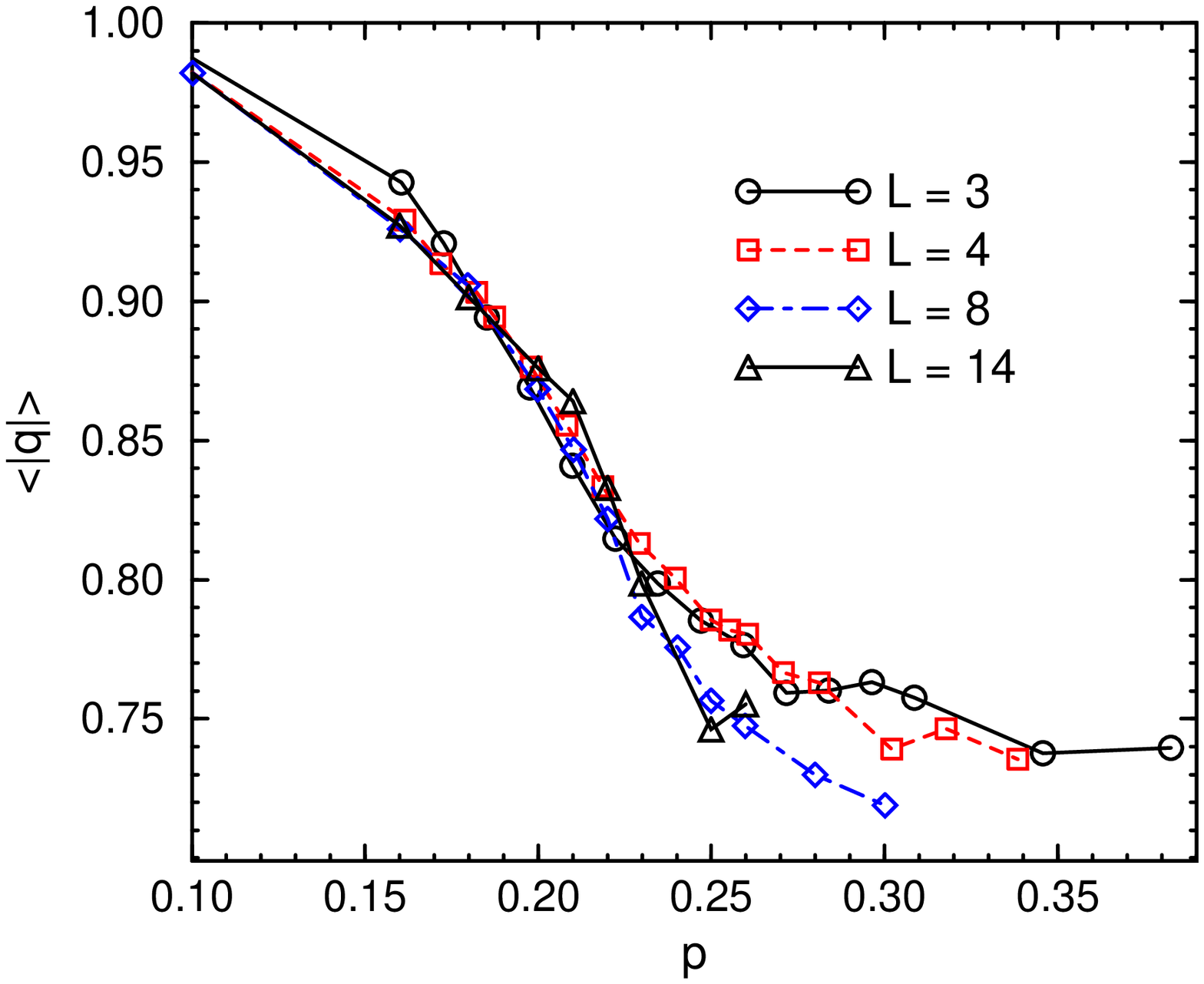}}
\end{center}
\caption{\captionE}
\label{fig_av_q}
\end{figure}
}

\newcommand{\figF}{
\begin{figure}[htb]
\begin{center}
\myscalebox{\includegraphics{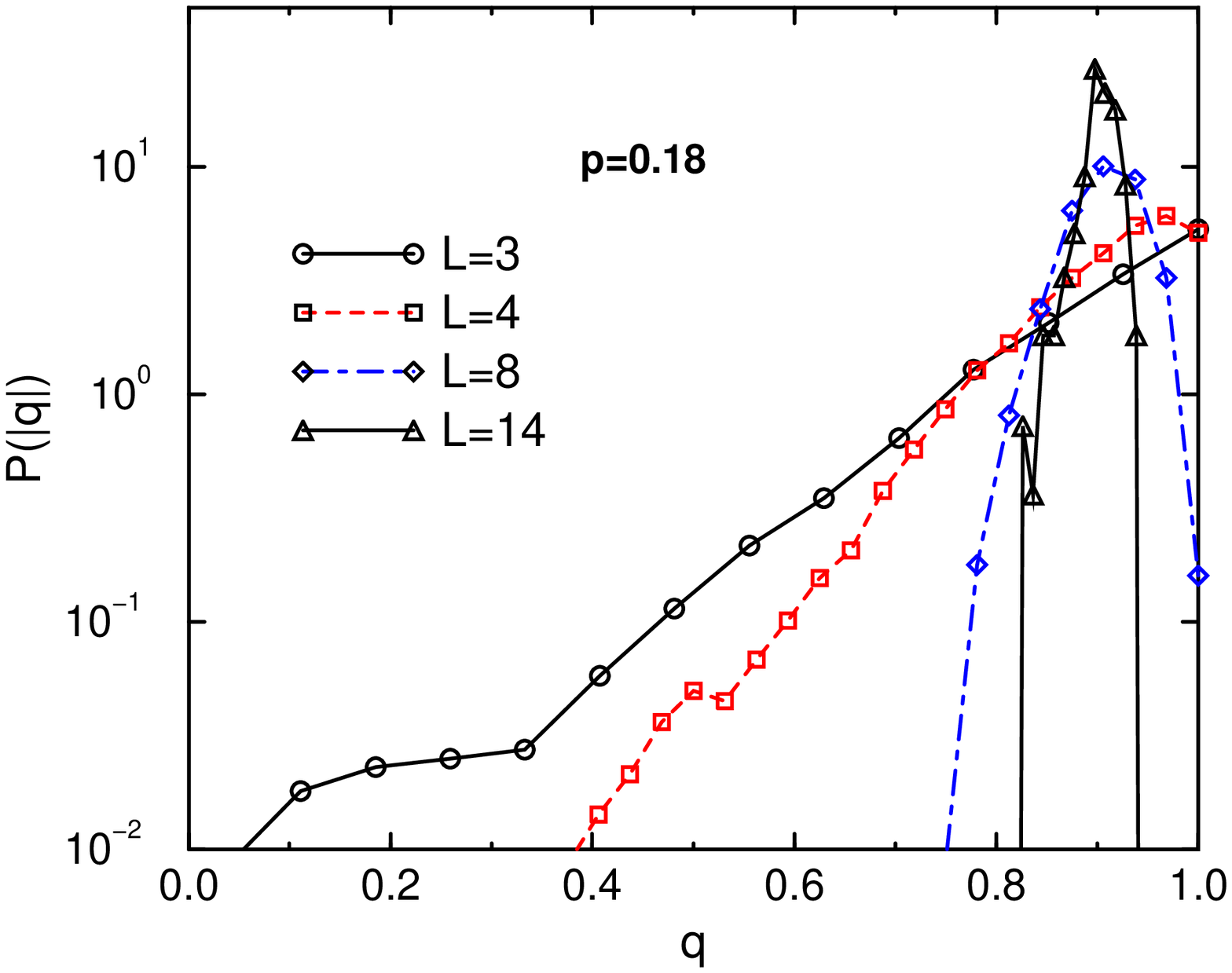}}
\end{center}
\caption{\captionF}
\label{fig_P_q_p0.18}
\end{figure}
}

\newcommand{\figG}{
\begin{figure}[htb]
\begin{center}
\myscalebox{\includegraphics{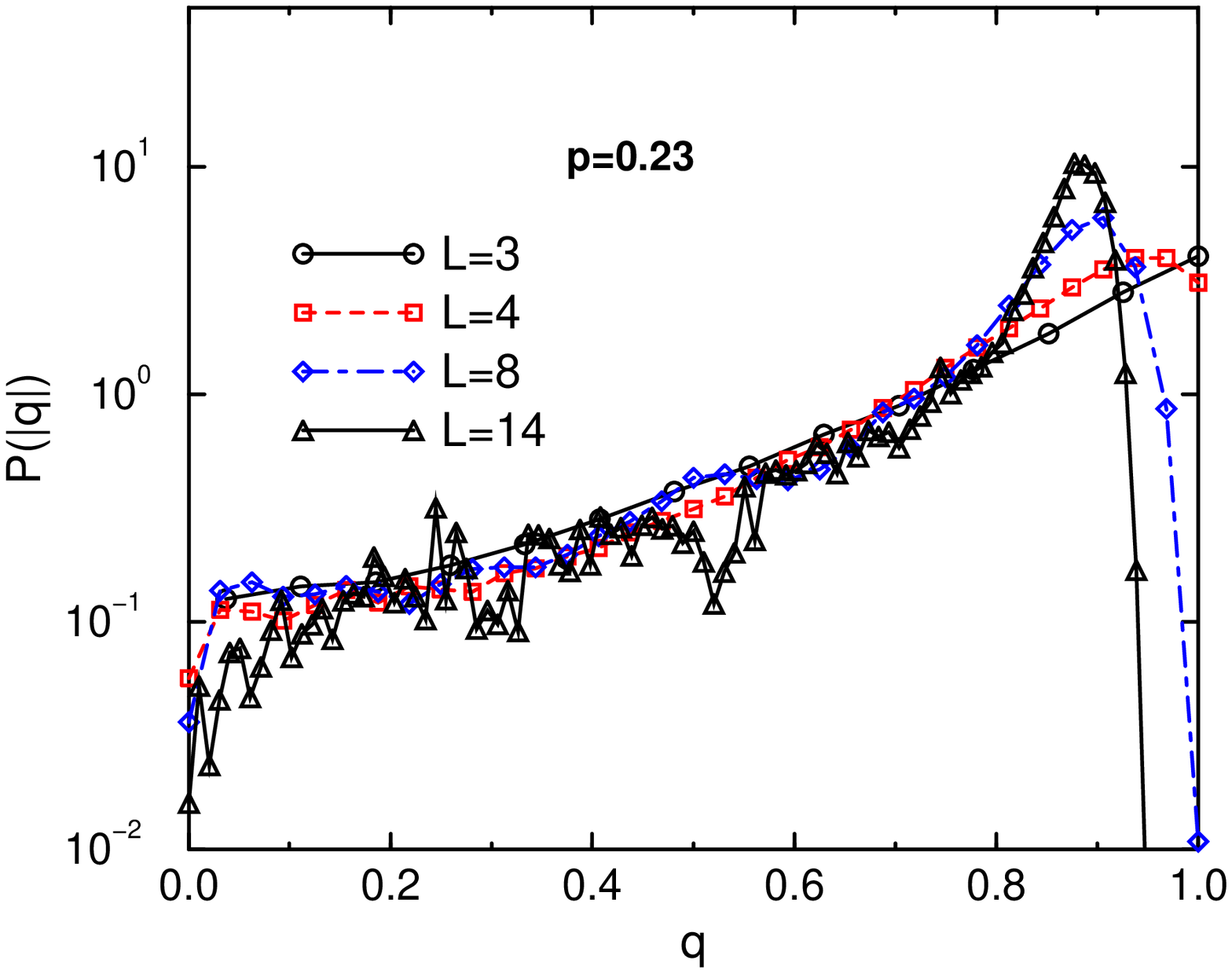}}
\end{center}
\caption{\captionG}
\label{fig_P_q_p0.23}
\end{figure}
}

\newcommand{\figH}{
\begin{figure}[htb]
\begin{center}
\myscalebox{\includegraphics{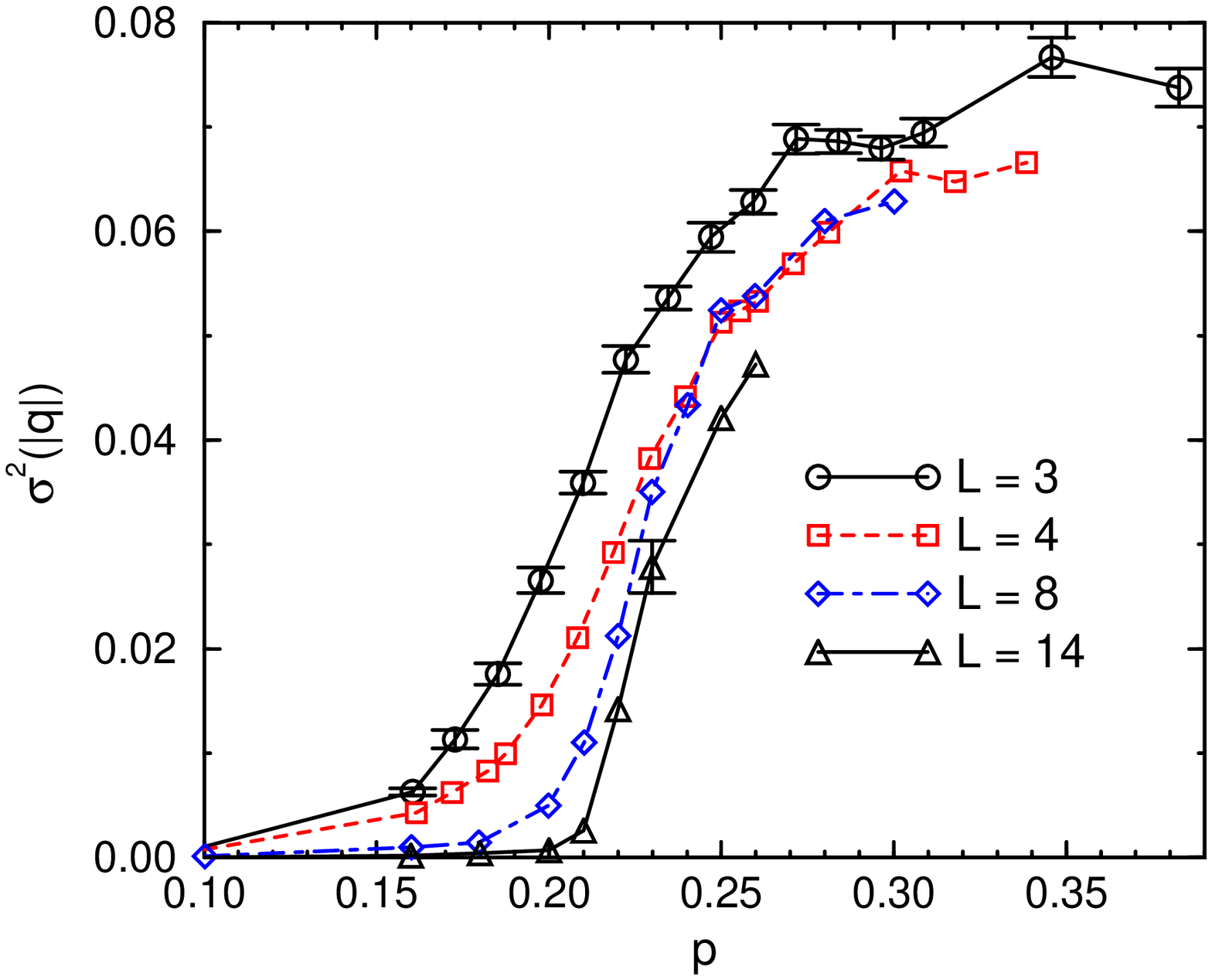}}
\end{center}
\caption{\captionH}
\label{fig_sigma_q}
\end{figure}
}

\newcommand{\figI}{
\begin{figure}[htb]
\begin{center}
\myscalebox{\includegraphics{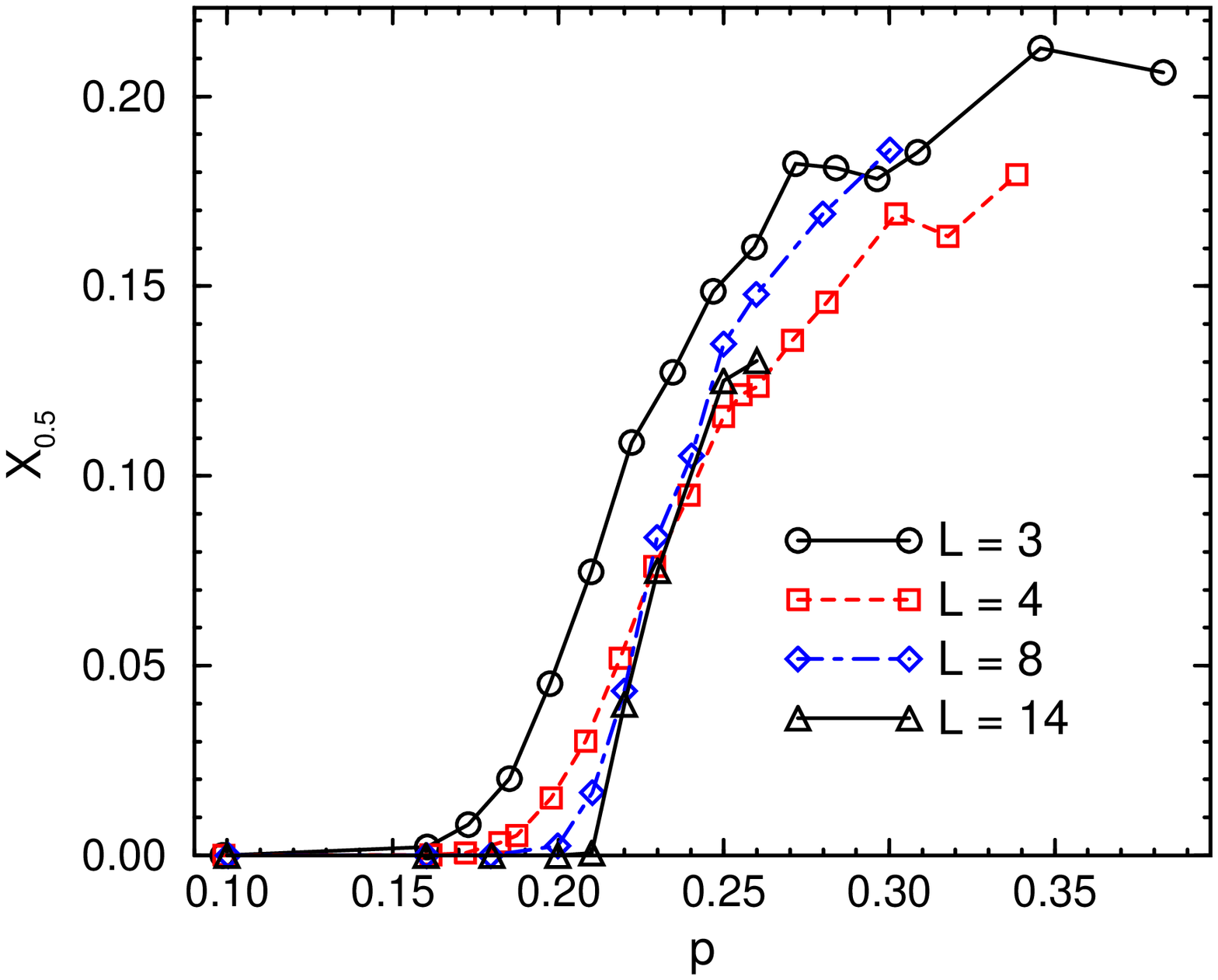}}
\end{center}
\caption{\captionI}
\label{fig_x05}
\end{figure}
}

\newcommand{\figK}{
\begin{figure}[htb]
\begin{center}
\myscaleboxc{\includegraphics{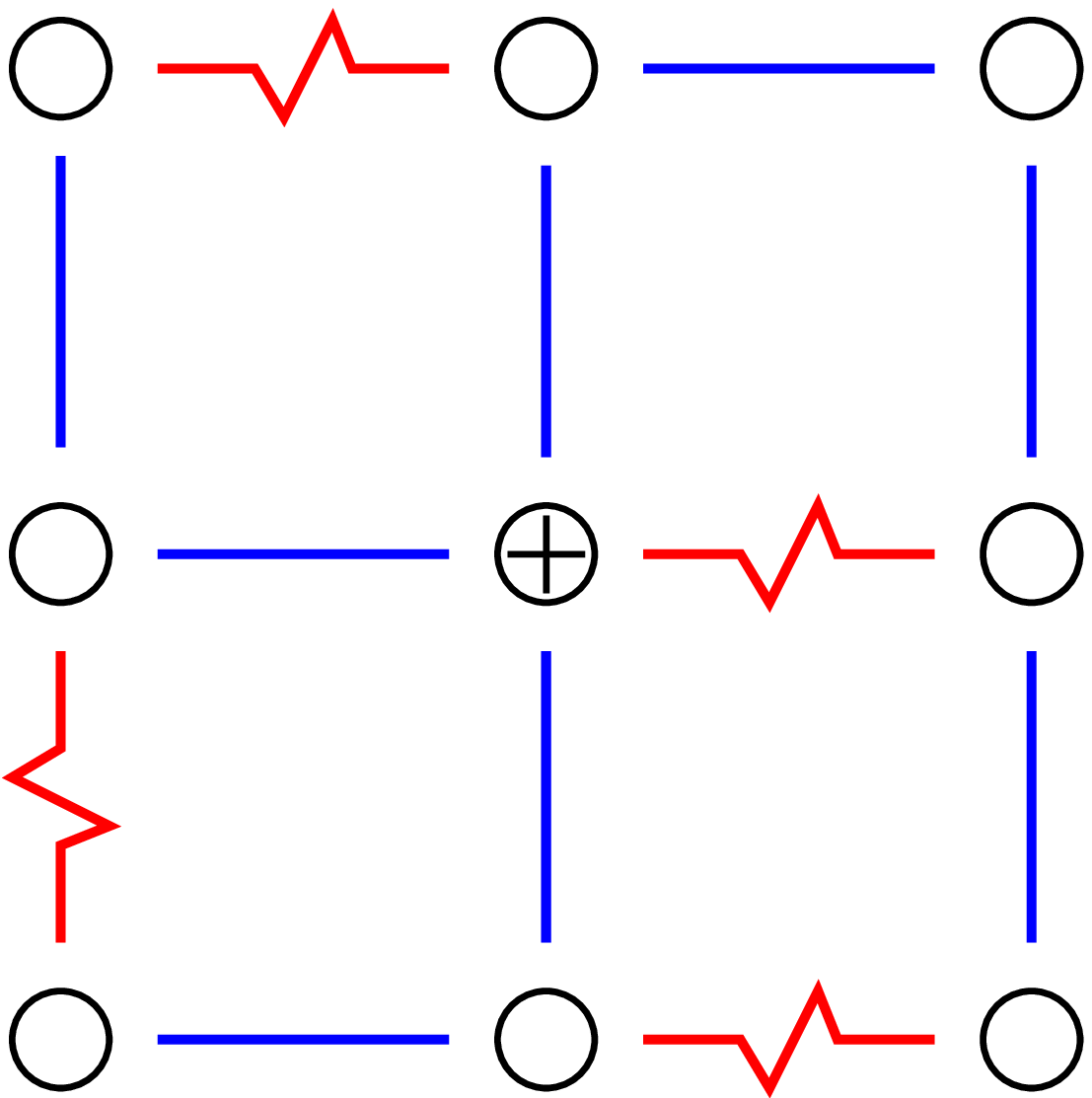}}

\vspace{0.2cm}

\myscaleboxc{\includegraphics{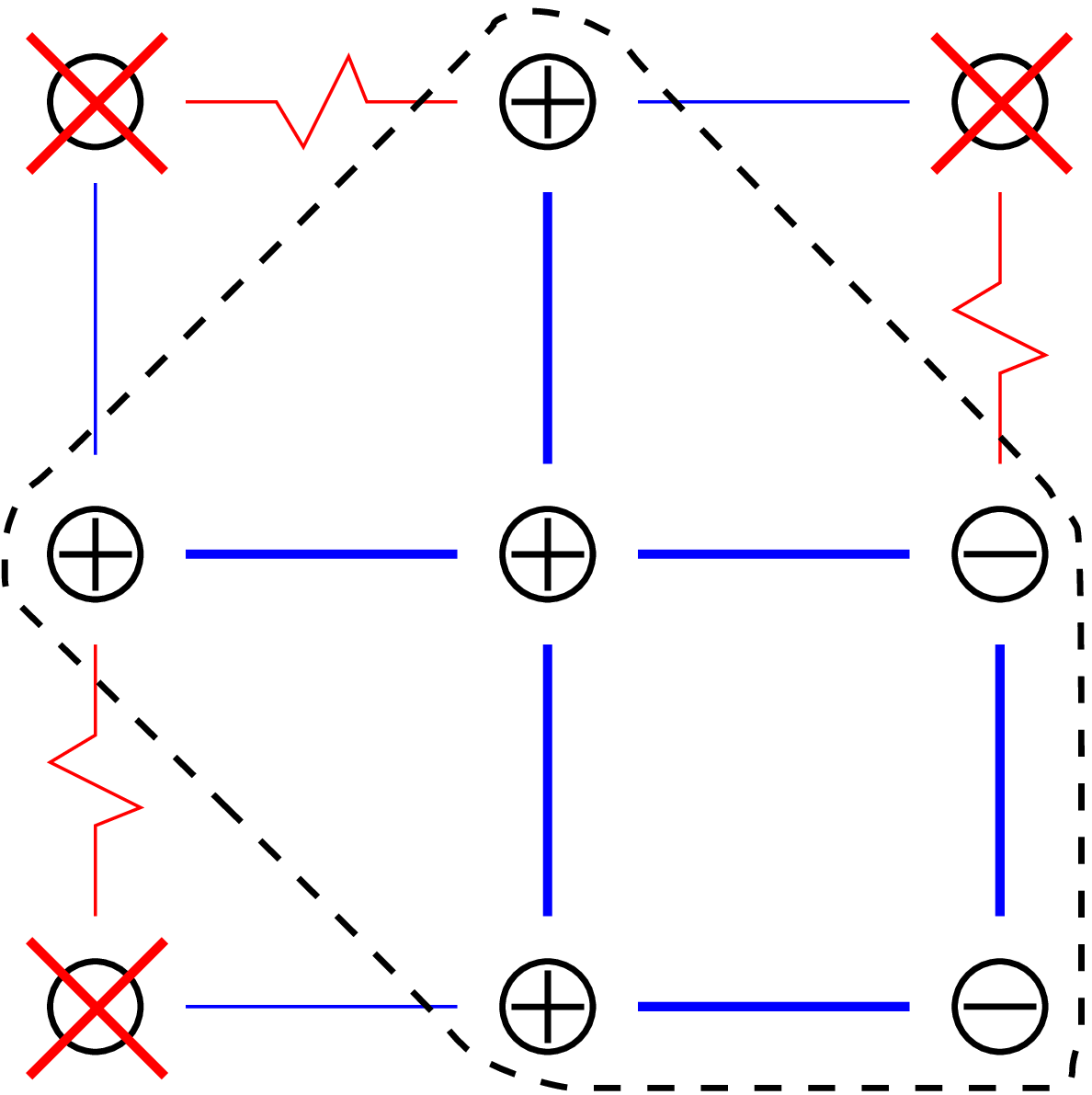}}
\end{center}
\caption{\captionK}
\label{fig_cea_example}
\end{figure}
}

\newcommand{\tabA}
{
\begin{center}
\begin{tabular}{ccccc}
\hline
$L$ & $M_i$ & $n_R$ & $n_{\min}$ & $\tau$ (sec) \\ \hline
3 & 16 & 3 & 1 & 0.2 \\
4 & 16 & 3 & 1 & 0.5 \\
5 & 16 & 4 & 2 & 3 \\
6 & 16 & 4 & 2 & 5 \\
8 & 32 & 4 & 5 & 70 \\
10 & 64 & 6 & 10 & 960  \\
14 & 256 & 14 & 10 & 32400
\end{tabular}
\end{center}

\vspace{0.2cm}

{Tab 1. Simulation parameters: $L$ = system size, $M_i$ = initial size of
population, $n_R$ = average number of offsprings per configuration, $n_{\min}$
= number of CEA minimization steps per offspring, $\tau$ = average computer
time per ground state on a 80MHz PPC601.}

\vspace{0.2cm} 

}

\begin{document}
\title{Ground-state behavior of the 3d $\pm J$ random-bond Ising model}

\author{Alexander K. Hartmann\\
{\small  hartmann@tphys.uni-heidelberg.de}\\
{\small  Institut f\"ur theoretische Physik, Philosophenweg 19, }\\
{\small 69120 Heidelberg, Germany}\\
{\small Tel. +49-6221-549449, Fax. +49-6221-549331}}

\date{\today}
\maketitle
\begin{abstract}
Large numbers of ground states of the three-dimensional
$\pm J$ random-bond Ising model are calculated
for sizes up to $14^3$ using a combination 
of a genetic algorithm and Cluster-Exact
Approximation. Several quantities are calculated as function of 
the concentration $p$ of the antiferromagnetic bonds. The critical
concentration where the ferromagnetic order disappears is
determined using the Binder cumulant of the magnetization. A value
of $p_c=0.222\pm 0.005$ is obtained. From the finite-size behavior
 of the Binder
cumulant and the magnetization critical exponents $\nu=1.1 \pm 0.3$
and $\beta=0.2 \pm 0.1$ are calculated. 

The behavior of the distribution
of overlaps $P(q)$ is used to investigate how the spin-glass phase evolves
with increasing concentration $p$. The spin-glass order is characterized
by a broad distribution of overlaps which extends down to $q=0$.

{\bf Keywords (PACS-codes)}: Spin glasses and other random models (75.10.Nr), 
Numerical simulation studies (75.40.Mg),
General mathematical systems (02.10.Jf). 
\end{abstract}


\paragraph*{Introduction}

In this work systems of $N$ spins 
$\sigma_i = \pm 1$, described by the Hamiltonian
\begin{equation}
H \equiv - \sum_{\langle i,j\rangle} J_{ij} \sigma_i \sigma_j
\end{equation}
are  investigated.
The spins are placed on a
three-dimensional (d=3) 
 cubic lattice of linear size $L$ with periodic boundary conditions in
all directions.
Systems with quenched disorder of the nearest-neighbor interactions (bonds)
are investigated. Their possible values are $J_{ij}=\pm 1$.
The concentration of the
antiferromagnetic (AF) bonds ($J_{ij}=-1$) is denoted with $p$, all other
$(1-p)$ interactions are ferromagnetic. A
constrained disorder is used, so that the fraction of the
antiferromagnetic bonds is exactly $p$ for all realizations of
the disorder.

The model shows a complex behavior for low temperatures.
For large concentrations of the ferromagnetic bonds it is
energetically favorable for two interacting spins to have the
same orientation. So the system shows ferromagnetic order, which
means that most of the spins have the same value. For
large concentrations $p$ the system is antiferromagnetically ordered:
the system can be divided into to penetrating sublattices and
each sublattice has ferromagnetic order, but the sign of the
ordering of the two sublattices is different.
For intermediate concentrations of the antiferromagnetic bonds neither
ferromagnetic nor antiferromagnetic order exists. The system
is called a spin glass \cite{binder86}. For finite-dimensional
spin glasses no final agreement about their behavior exists. 
Recent results from simulations\cite{marinari96,alex_sg2} of
small systems with $p=0.5$ up to size $16^3$ 
indicate that the three-dimensional spin glass has a complex behavior:
the (free) energy landscape has many stable configurations which differ
strongly from each other. Whether the onset of this spin-glass behavior
takes place at the same concentration where the ferromagnetic order
disappears is unclear.

Here the ground-state, i.e. zero temperature ($T=0$), behavior 
 of the model as function of the concentration $p$ 
is investigated. Since the phase diagram of the
system is symmetrical to $p=0.5$ if one identifies the ferromagnetic
with the antiferromagnetic regime, only values $p<0.5$ are used.
Of special interest is the
critical concentration $p_c$ where the ferromagnetic order disappears. 

In the past the $\pm J$ random-bond Ising model has been studied using 
Monte-Carlo simulations\cite{kirkpatrick77},
zero-temperature series expansions \cite{grinstein79}, 
high-temperature series expansions \cite{reger86,fisch91},
Monte Carlo renormalization-group calculations \cite{ozeki87} and
renormalization-group theory \cite{migliorini98}. All of these
results are qualitatively consistent with the phase-diagram
described above, but no agreement is found
 on the value of the critical concentration $p_c$ or
its temperature dependence. In all of  these
publications the detailed 
structure of the states was not investigated, which can
be done for example by calculating the distribution of overlaps.

A study of this model using true ground states has not been performed
before. Only for special two-dimensional systems, where exact ground states
can be calculated efficiently, some results 
\cite{bendisch92,bendisch94,kawashima97} 
are known. There the critical concentration of a square lattice
is estimated to be $p_c^{2d}=0.10$. 

The behavior of the $\pm J$ random bond Ising model
 is determined by the occurrence of {\em frustration} \cite{toulouse77}. 
The simplest example of a frustrated system is a triple
of spins where all pairs are connected by antiferromagnetic bonds. It is
not possible to find a spin-configuration were all bonds contribute
with a negative value to the energy. One says that it is not possible
to {\em satisfy} all bonds. In general a system is frustrated, if closed
loops of bonds exists, where the product of these bond-values is negative.
For square and cubic systems the smallest closed loops consist of four
bonds. They are called (elementary) {\em plaquettes}.

The presence of frustration makes the calculation of exact ground states
of such systems computationally hard.
Only for the special case of the two-dimensional system with
periodic boundary conditions in no more than one direction and without
external field a polynomial-time algorithm is known
\cite{barahona82b}.  For the general case the simplest method works by
enumerating all $2^N$ possible states and has obviously an exponential time
complexity. Even a system size of $4^3$ is too large.  The basic idea 
of an algorithm
called {\em Branch-and-Bound} \cite{hartwig84} is to exclude
the branches of the search-tree, where no low-lying states can be found, so
that the complete low-energy landscape of systems of size $4^3$ 
can be calculated \cite{klotz}. A more
sophisticated method called {\em Branch-and-Cut} \cite{simone95,simone96}
works by rewriting the problem as a linear optimization problem with
an additional set of inequalities which must hold for the solution.
Since not all inequalities are known a priori the method iteratively
solves the linear problem, looks for inequalities which are violated,
and adds them to the set until the solution is found. Since the number
of inequalities grows exponentially with the system size the same
holds for the computation time of the algorithm. With Branch-and-Cut
anyway small systems
up to $8^3$ are feasible. The method used here is able to
calculate true ground states up to size $14^3$.

By studying ground
states one does not encounter ergodicity problems or critical
slowing down like in Monte-Carlo simulations. Since it is
possible to compare with exact results from Branch-and-Cut calculations
no uncontrolled approximations are used. The only uncertainty comes from
the fact that one is restricted to relatively small systems.

In the next section the algorithm is explained. Then all results
are presented. In the last section a conclusion is driven.
 
\section*{Algorithm}
The algorithm for the calculation of the ground states 
bases on a special genetic
algorithm \cite{pal96,michal92} and on {\em Cluster-Exact Approximation}  
(CEA) \cite{alex2} which is  a sophisticated optimization method.
Next a short sketch of these algorithms is given.

The genetic algorithm starts with an initial population of $M_i$
randomly initialized spin configurations (= {\em individuals}),
which are linearly arranged in
a ring. Then $n_R \times M_i$ times two neighbors from the population
are taken (called {\em parents}) and two offsprings are created
using a triadic crossover: a mask is used which is a
third randomly chosen (usually distant) member of the population with
a fraction of $0.1$ of its spins reversed. In a first step the
offsprings are created as copies of the parents. Then those spins are selected,
 where the orientations of the
first parent and the mask agree \cite{pal95}. 
The values of these spins
are swapped between the two offsprings. Then a {\em mutation}
 with a rate of $p_m$
is applied to each offspring, i.e. a fraction $p_m$ of the
spins is reversed.

Next for both offsprings the energy is reduced by applying
CEA:
The method constructs iteratively and randomly 
a non-frustrated cluster of spins.
Spins adjacent to many unsatisfied bonds are more likely to be added to the
cluster. During the construction of the cluster a local gauge-transformation
of the spin variables is applied so that all interactions between cluster
spins become ferromagnetic.

\figK

Fig. \ref{fig_cea_example} shows an example of how the construction 
of the cluster works using a small spin-glass system.
For 3d $\pm J$ spin glasses each cluster
contains typically 58 percent of all spins.
The  non-cluster spins act like local magnetic fields on the cluster spins,
so the ground state of the cluster is not trivial.
Since the cluster has only ferromagnetic interactions, 
an energetic minimum state for its spins can be  calculated in polynomial time
by using graph theoretical methods 
\cite{claibo,knoedel,swamy}: an equivalent network is constructed
\cite{picard1}, the maximum flow is calculated 
\cite{traeff,tarjan}\footnote{Implementation details: We used 
Tarjan's wave algorithm together
with the heuristic speed-ups of Tr\"aff. In the construction of 
the {\em level graph} we allowed not only edges $(v, w)$
with level($w$) = level($v$)+1, but also all edges $(v,t)$ where $t$
is the sink. For this measure, we observed an additional speed-up of
roughly factor 2 for the systems we calculated.} and the spins of the
cluster are set to their orientations leading to a minimum in energy. 
This minimization step
is performed $n_{\min}$ times for each offspring.

Afterwards each offspring is compared with one of its parents. The
pairs are chosen in the way that the sum of the phenotypic differences
between them is minimal. The phenotypic difference is defined here as the
number of spins where the two configurations differ. Each
parent is replaced if its energy is not lower (i.e. not better) than the 
corresponding offspring.
After this whole step is done $n_R \times M_i$ times, the population
is halved: From each pair of neighbors the configuration 
 which has the higher energy is eliminated. If more than 4
individuals remain the process is continued otherwise it
is stopped and the best individual
is taken as result of the calculation.

The representation in fig. \ref{fig_algo} summarizes the algorithm. 

The whole algorithm is performed $n_R$ times and all configurations
which exhibit the lowest energy are stored, resulting in $n_G$ statistically
independent ground-state configurations.

This algorithm was already applied to examine the ground state 
structure of 3d spin glasses \cite{alex_sg2}.

\newlength{\mpwidth}
\setlength{\mpwidth}{\textwidth}
\addtolength{\mpwidth}{-2cm}
\begin{figure}

\begin{center}

\begin{minipage}[b]{\mpwidth}
\newlength{\tablen}
\settowidth{\tablen}{xxx}
\newcommand{\tabspace}{\hspace*{\tablen}}
\begin{tabbing}
\tabspace \= \tabspace \= \tabspace \= \tabspace \= \tabspace \=
\tabspace \= \kill
{\bf algorithm} genetic CEA($\{J_{ij}\}$,
$M_i$, $n_R$, $p_m$, $n_{\min}$)\\
{\bf begin}\\
\> create $M_i$ configurations randomly\\
\> {\bf while} ($M_i > 4$) {\bf do}\\
\> {\bf begin}\\
\> \> {\bf for} $i=1$ {\bf to} $n_R \times M_i$ {\bf do}\\
\>\> {\bf begin}\\
\>\>\> select two neighbors \\
\>\>\> create two offsprings using triadic crossover\\
\>\>\> do mutations with rate $p_m$\\
\>\>\> {\bf for} both offsprings {\bf do}\\
\>\>\> {\bf begin}\\
\>\>\>\> {\bf for} $j=1$ {\bf to} $n_{\min}$ {\bf do}\\
\>\>\>\> {\bf begin}\\
\>\>\>\>\> construct unfrustrated cluster of spins\\
\>\>\>\>\> construct equivalent network\\
\>\>\>\>\> calculate maximum flow\\
\>\>\>\>\> construct minimum cut\\
\>\>\>\>\> set new orientations of cluster spins\\
\>\>\>\> {\bf end}\\
\>\>\>\> {\bf if} offspring is not worse than related parent \\
\>\>\>\> {\bf then}\\
\>\>\>\>\> replace parent with offspring\\
\>\>\> {\bf end}\\
\>\> {\bf end}\\
\>\> half population; $M_i=M_i/2$\\
\> {\bf end}\\
\> {\bf return} one configuration with lowest energy\\
{\bf end}
\end{tabbing}

\end{minipage}
\end{center}
\caption{Genetic Cluster-exact Approximation.}
\label{fig_algo}
\vspace{0.5cm}
\end{figure}

\section*{Results}
We used the simulation parameters determined in  former calculations for
$p=0.5$:
For each system size  many different combinations
of the simulation
parameters $m_i, n_R, n_{min}, p_m$  were tried for some sample systems. 
The final
parameters where determined in a way, that
by using four times the numerical effort no reduction in
energy was obtained. Here $p_m=0.2$ and $n_R=10$ were used
for all system sizes.

For smaller concentrations $p$ the ground states are easier to find,
because the number of frustrated plaquettes is smaller. But it
was not possible to reduce the computational effort substantially
in order to get still ground states.
So we used the parameters of $p=0.5$ for  all concentrations $p$.
Table 1 summarizes the parameters. Also the
typical computer time $\tau$ per ground state
computation on a 80 MHz PPC601 is given. 
Using these parameters on average $n_G>8$
ground states were obtained for every system size $L$ using  $n_R=10$ runs per
realization.

\tabA

 We compared our results for 180 sample systems of $L=6$ 
 with exact ground states which were obtained
using a  Branch-and-Cut
program \cite{simone95,simone96}. 
The genetic CEA algorithm found the true ground states for all systems!
The same result was obtained for $L=4$ as well.\footnote{For $L>6$ the 
Branch-and-Cut program needs to much computer time because of the exponential
time complexity.} A more detailed analysis is presented
in \cite{alex_stiff}. So we can be sure that genetic CEA
and our method of choosing the parameters lead to true ground states or
at least to states very close to true ground states.

We performed ground state calculations for $p\in[0.1,0.4]$ for
lattice sizes $L=3,4,5,6,8,10,14$. The number of independent
realizations of the bond-disorder ranged from $N_L=1000$ for $L=14$ to
$30000$ for smaller systems. Most of the systems have AF-bond concentrations
around $p=0.22$

\figA

The ground state energy $e$ as function of system size for
 different system sizes
is shown in Fig \ref{fig_energy}. To keep the figure clear only the sizes
$L=3,4,6,14$ are presented. For small concentrations $p$ 
the ground state is mainly ferromagnetic.
It follows that all AF bonds are not satisfied, so the ground state
energy increases linearly like $e(p)\approx -3+6p$ with $p$. 
For larger concentrations the
ground state energy approaches the $p=0.5$ limit, because the spins
can arrange so that not all AF-bonds are broken. 
With increasing $L$ the ground state energy decreases,
because the periodic boundary conditions impose less
constraints on the system. For $L\to\infty$ and $p=0.5$ a ground
state energy of $e_{\infty}(0.5)=-1.7876(3)$ is found in \cite{alex_sg2}.

\figB

From fig. \ref{fig_energy} is clear that the
 energy as function of concentration is not well suited for determining
the critical concentration $p_c$ 
where the ferromagnetic behavior disappears.
For this purpose the Binder cumulant \cite{binder81,bhatt85}

\begin{equation}
q(p,L)\equiv\frac{1}{2}
\left( 3-\frac{\langle M^4\rangle }{\langle M^2\rangle^2}\right)
\end{equation}

for the magnetization $M\equiv\frac{1}{N}\sum_i \sigma_i$ is used.
The average $\langle \ldots \rangle$ denotes both average over
different ground states of a realization and over the disorder.
In fig. \ref{fig_binder} the Binder cumulant is shown for $L=3,5,8,10,14$. 
L=$4,6$ are omitted in this figure
to keep it clear. For the same reason only typical error bars for two
sample points are shown. All curves intersect at $p_c=0.222\pm 0.002$.
Only $L=4$ (not shown) is little worse, because it meets the others in the
interval $p\in[0.217,0.222]$. So we conclude that the critical concentration
for the ferromagnetic order is $p_c=0.222(5)$.

The value for $p_c$ is comparable to results from high-temperature
series expansions: $p_c=0.19(2)$\cite{reger86}, $p_c\approx 0.25$ 
\cite{fisch91}, from Monte-Carlo renormalization-group results
$p_c=0.233(4)$ \cite{ozeki87} 
and from Monte-Carlo simulations: $p_c\approx 0.24$
\cite{kirkpatrick77}. Our value is much larger than a result
from a zero-temperature expansion: $p_c=0.12-0.13$ \cite{grinstein79}
and much lower than a recent result from a renormalization-group study:
$p_c\approx 0.37$ \cite{migliorini98}.
Since the
intersection of the curves of the Binder cumulant is very sharp, we
believe that our result is very reliable, 
although the systems investigated here are rather small.

\figC

For the Binder cumulant the following finite-size scaling relation
is assumed \cite{bhatt85}

\begin{equation}
q(p,L)=\tilde{g}(L^{1/\nu}(p-p_c))
\end{equation}

By plotting $g(p,L)$ against $L^{1/\nu}(p-p_c)$ with correct
parameter $\nu$ the datapoints for different system sizes should
collapse onto a single curve. The best results were obtained for
$p_c=0.222$ and $1/\nu=0.9$. In fig. \ref{fig_binder_scale} the resulting
scaling plot is shown. It is possible to change the value of $\nu$ in
a wide range without large effects on the scaling plot. So we
estimate $\nu=1.1(3)$. This is 
consistent with $\nu=1.7(3)$ which was found using Monte-Carlo simulations
of spin glasses ($p=0.5$) at finite temperature \cite{kawashima96}.

The average magnetization $m\equiv\langle M \rangle$ has 
the standard finite-size  scaling form \cite{binder_heermann}
\begin{equation}
m(p,L)=L^{-\beta/\nu}\tilde{m}(L^{1/\nu}(p-p_c))
\end{equation}

By plotting $L^{\beta/\nu}m(p,L)$ against $L^{1/\nu}(p-p_c)$ with correct
parameters $\beta,\nu$ the datapoints for different system sizes should
collapse onto a single curve.
The best result was obtained using $1/\nu=0.9$ and $\beta/\nu=0.19$.
It is shown in fig. \ref{fig_magn_scale} for $L=3,5,8,10,14$.
From variations of the value $\beta/\nu$ we estimate the
value of the exponent $\beta=0.2(1)$.

\figD

To characterize the spin-glass behavior the overlap $q$ is used. It
compares two different states $\{\sigma_i^{\alpha}\}, \{\sigma_i^{\beta}\}$ 
of the same realization of the random bonds

\begin{equation}
q^{\alpha\beta} \equiv \frac{1}{N}\sum_i \sigma_i^{\alpha} \sigma_i^{\beta}
\end{equation}

For ferromagnetic/antiferromagnetic order two independently calculated
ground states are identical or related by a global flip of all spins, i.e.
$q=\pm1$. For spin-glass order many different ground states exist 
\cite{alex_sg2}, so $q$ can take also intermediate values $q\in [-1,1]$.
Since the system has no external field, each state is equivalent to the
state where all spins are reversed, so only the absolute value $|q|$ is 
considered here.

\figE

In fig. \ref{fig_av_q} the average value of the overlap 
$\langle |q| \rangle \equiv \langle |q^{\alpha\beta}|\rangle$
is shown
for the lattice sizes $L=3,4,8,14$. 
The decrease of $\langle |q| \rangle$
with increasing concentration $p$ is clearly visible. But this quantity has
much larger fluctuations than the magnetization, so it is 
difficult to use this data for further analysis. Also the value is
not equal to $1.0$ for a large interval, so it is not possible to
extract at which concentration the onset of the spin-glass behavior 
is located.

\figF

More information can be obtained, if one calculates not only the
average of $|q|$, but its distribution \cite{parisi2}

\begin{equation}
P(|q|)\equiv \langle \delta(|q|-|q^{\alpha\beta}|) \rangle
\end{equation}
 which describes the ground-state structure. In 
fig. \ref{fig_P_q_p0.18}\
the distributions for sizes $L=3,4,8,14$ at $p\approx 0.18$ are 
presented\footnote{Since all realizations of a given size $L$ have
exactly the same number of ferromagnetic bonds, for each system size
only a finite number of different concentrations is possible. 
Here for each lattice size the value of $p$ is chosen which 
is nearest to $0.18$}.
With increasing size the distributions becomes narrower. The reason is,
that due to the discrete bond distribution $J_{ij}=\pm 1$ there
are always some small clusters of spins, which can take two orientations
in the ground state. With increasing system size $L$ these effects cancel
out and $P(|q|)$ converges to a delta-function $\delta(q-q_{free})$, where
$q_{free}$ is just the fraction of spins contained in such free clusters.
This ground-state structure is similar to that of random-field Ising
systems \cite{alex_daff2}.

\figG

In fig. \ref{fig_P_q_p0.23} the distributions of overlaps is
displayed for a concentration slightly larger than $p_c$. Here the
behavior is completely different. The distributions extend over a large
interval down to $q=0$. With increasing lattice size $L$ only the
shape of the large-q part changes a little bit while for small values
no systematic modification is visible. The peak at large $q$-values, which
raises with increasing concentration $p$, results from small clusters of
spins which can take two orientations in the ground state. This
is the same as for  smaller concentrations of the AF-bonds. 
But the large extent
down to $q=0$ cannot be explained in this way. It shows that spin-glass
ground states have a very rich structure, similar to the behavior found
for the SK-model \cite{parisi2}, 
where each spin interacts with every other spin (for a detailed discussion of
the ground-state structure see \cite{alex_sg2,alex_ultra}).

To investigate this behavior quantitatively the variance $\sigma^2(|q|)$ of the
distributions as function of AF-bond concentration $p$ and system
size $L$ is calculated

\begin{equation}
\sigma^2(|q|)\equiv \langle (|q^{\alpha\beta}|-\langle |q| \rangle )^2\rangle
\end{equation}

\figH

In fig. \ref{fig_sigma_q} the result for $L=3,4,8,14$ is shown. For small
concentration $p$ the width of the distributions is small and shrinks
with increasing size $L$. For larger concentrations the width increases
and remains nonzero even for larger lattice sizes. In \cite{alex_sg2} it
was shown, that the spin glass ground state 
with $p=0.5$ is likely to have a broad distribution even for $L\to\infty$.
For larger sizes the statistics is too bad to extract for
example a critical concentration $p_c^{(2)}$ by a finite-size scaling analysis
similar to that presented above.

The onset of the spin glass behavior can even better be observed by
calculating the fraction $X_{q_0}$ of the distribution of overlaps
below a fixed value $q_0$:

\begin{equation}
X_{q_0}\equiv \int_0^{q_0} P(|q|)\,dq
\end{equation}

\figI

In fig. \ref{fig_x05} the value of $X_{0.5}$ is shown as function of $p$
for the lattice sizes $L=3,4,8,14$. For the limiting case $p=0.5$ 
the value of $X_{0.5}$ converges to a nonzero value for $L\to\infty$
\cite{alex_x05}.

\section*{Conclusion}

Using a combination of a genetic algorithm and Cluster-Exact Approximation
ground states of the three-dimensional $\pm J$ random-bond Ising model were
calculated for different concentrations of the antiferromagnetic bonds.
A former comparison with exact ground states calculated using 
a Branch-and Cut program
shows that genetic CEA is able to calculate true ground states.

For small concentrations the ground state is mainly ferromagnetic. The
critical concentration where the ferromagnetic order disappears was
determined using the Binder-cumulant $g(p,L)$ 
of the magnetization: 

\begin{equation}
p_c=0.222 \pm 0.005
\end{equation}

 This is the first time $p_c$
is calculated by direct investigations of ground states.

Using a finite-size scaling analysis of the magnetization and the
Binder-cumulant critical exponents were obtained: 

\begin{equation}
\nu=1.1\pm 0.3,\,\beta=0.2\pm 0.1
\end{equation}

 These values are consistent with results from Monte-Carlo
simulations of the $p=0.5$ case at finite temperature, where the same analysis
presented here was performed for the spin-glass order parameter $q$ instead
of the magnetization $m$.

The onset of the spin-glass behavior with increasing lattice size was
investigated the first time 
by calculating the distributions of overlaps as function of $p$. The spin
glass phase is characterized by a broad distribution of overlaps
which extends down to $q=0$ and does not change substantially with
increasing system size. For this quantity the bad statistics for larger
values of the concentration $p>0.23$ does not
allow to determine a second critical concentration $p_c^{(2)}$, so it
is yet not possible to check whether the ferromagnetic
order disappears at the same concentration where the spin-glass
phase appears. Here much more data are needed.

\section*{Acknowledgements}

The author thanks H. Horner and G. Reinelt for manifold support.
He is grateful to M. J\"unger, M. Diehl and T. Christof who put
a Branch-and-Cut program for the exact calculation of spin-glass
ground states of small systems at his disposal.
He thanks R. K\"uhn for critical reading of the manuscript and for giving 
many helpful hints.
This work was engendered by interesting discussions with 
J. Bendisch at the 
``Algorithmic Techniques in Physics'' seminar held
at the {\em International Conference and Research Center for
Computer Science Schloss Dagstuhl} in Wadern/Germany.
The author took much benefit from discussions with S. Kobe, H. Rieger
and A.P. Young.
He thanks F. Hucht for supplying the program {\em fsscale} for
performing the finite-size scaling analysis.
He is also grateful to the {\em Paderborn Center for Parallel Computing}
 for the allocation of computer time. This work was supported
by the Graduiertenkolleg ``Modellierung und Wissenschaftliches Rechnen in 
Mathematik und Naturwissenschaften'' at the
{\em In\-ter\-diszi\-pli\-n\"a\-res Zentrum f\"ur Wissenschaftliches Rechnen}
 in Heidelberg.





\end{document}